\documentclass[12pt,preprint]{aastex}

\shorttitle{Search for the Missing Baryons}
\shortauthors{Bregman et al.}
\usepackage{amstext}

\begin{document}

\title{X-Ray Searches for Emission from the WHIM in the Galactic Halo and the Intergalactic Medium}

\author{Joel N. Bregman, Birgit Otte, Jimmy A. Irwin, Mary E. Putman}
\affil{Department of Astronomy, University of Michigan, Ann Arbor, MI 48109}
\author{Edward J. Lloyd-Davies}
\affil{Astronomy Centre, Department of Physics and Astronomy, University of Sussex}
\author{and Christian Br\"uns}
\affil{Argelander-Institut f\"ur Astronomie, Universit\"at Bonn, Bonn, Germany}
\email{jbregman@umich.edu, otteb@umich.edu, jairwin@umich.edu,
mputman@astro.columbia.edu, E.Lloyd-Davies@sussex.ac.uk, cbruens@astro.uni-bonn.de}

\begin{abstract}

At least 50\% of the baryons in the local universe are undetected and 
predicted to be in a hot dilute phase (10$^{5}$-10$^{7}$ K) in low and moderate
overdensity environments.  We searched for the predicted diffuse faint emission 
through shadowing observations whereby cool foreground gas absorbs more distant 
diffuse emission.  Observations were obtained with {\it Chandra\/} and {\it XMM-Newton\/}.
Using the cold gas in two galaxies, NGC 891 and NGC 5907, shadows were not detected
and a newer observation of NGC 891 fails to confirm a previously reported X-ray shadow.
Our upper limits lie above model predictions.  For Local Group studies, we used a
cloud in the Magellanic Stream and a compact high velocity cloud to search for 
a shadow.  Instead of a shadow, the X-ray emission was brighter towards the Magellanic
Stream cloud and there is a less significant brightness enhancement toward the other cloud also.  
The brightness enhancement toward the Magellanic Stream cloud is probably
due to an interaction with a hot ambient medium that surrounds the Milky Way.  
We suggest that this interaction drives a shock into the cloud, heating the gas 
to X-ray emitting temperatures.

\end{abstract}


\section{Introduction}

An inventory of the baryons at the present epoch shows that the visible
galaxies and the hot gas in clusters and rich groups comprise only a
fraction of the baryons inferred from cosmic microwave background observations \citep{sperg07} or from the D/H ratio, combined with  Big Bang nucleosynthesis calculations \citep{kirk03}; these two values are in good agreement.  Most of
the baryons must be in a form that is difficult to detect, and cosmological
simulations suggest that it is still in the gaseous phase in regions of low
and moderate overdensities ($\rho/\bigtriangleup\rho = 10-100$; 
\citealt{fuku98,cen99,dave01, croft01,cen06a}).  
In the simulations, gravitational collapse leads to
compressional and shock heating of the gas, producing temperatures of
10$^{5}$ - 10$^{7}$ K, where hydrogen is almost completely ionized.  The metallicity of
this gas increases with cosmic time due to enrichment by galactic winds,
leading to present day metallicities of about 1/10 of the solar value \citep{danf05}.

At X-ray energies, one can try to detect this gas in emission or in
absorption and each approach poses substantial challenges.  For
absorption line studies, the intergalactic medium (IGM) is predicted to
produce only small equivalent width lines that are at the limit of
detectability (reviews by \citealt{breg07rev,nica08}).  Another approach is to search for
emission from the IGM, since the presence of metals, and especially of Fe and O,
produces X-ray line emission, which when summed over many filaments
and large volumes, should be detectable as a diffuse X-ray background
\citep{cen99,croft01}. 
This diffuse background is near the present limit of detectability and is
predicted to be less than the typical instrumental backgrounds or than the
point source background, even in the 0.4--1 keV range where IGM
emission is expected to be most prominent.  It is not possible to subtract
the unresolved point source background and the instrumental background
to sufficient accuracy, as they can vary from field to field and are not
known with sufficient accuracy a priori.

A detection method that avoids this problem is that of X-ray shadowing,
whereby an excess of cold gas (e.g., a gas cloud) lies in front of the
emitting source and falls within the field of view of the detector.  The
cloud absorbs some of the background X-ray emission, producing a local
minimum on the image.  To apply this method, it is crucial that the
distance to the absorbing cloud is known and that it's absorption column
has been measured independently, such as from \ion{H}{1} measurements.  This
shadowing technique has been applied within the Milky Way to show that
there is a hot halo above the disk of the Galaxy \citep{snow97} and that
there is X-ray emission from the Galactic bulge \citep{park97}.

Here we determine whether shadows are cast in two situations that
address the missing baryon problem.  The first set of observations uses
the \ion{H}{1} in external galaxies to shadow the background emission from the
ensemble of hot cosmic filaments.  This was utilized in the work of
\citet{breg02}, where the dense atomic and molecular gas in the disk of
NGC 891 appeared to produce a shadow at the 98\% confidence level. 
That is, in the 0.4--1.0 keV band, there were 39.5 photons in the
extinction region compared to 57.6 photons in the non-extinction region,
a deficit of 18 photons.  Since this is less than a 3$\sigma$ effect, it is
important
to confirm or refute the result, and as the initial observation was 51 ksec,
there is room for improvement.  A significantly longer observation (120
ksec) was obtained for another program, which permits us to examine the
quality of the shadow seen with the shorter observation.  A second
shadowing test is also presented, whereby we use the very extended \ion{H}{1}
from the edge-on galaxy NGC 5907 to search for a shadow against the
diffuse IGM emission.

The other location that one might hope to find the missing baryons is in the
Local Group, where the X-ray absorption lines from \ion{O}{7} and \ion{O}{8}
have been detected \citep{nica03,rasm03} as has OVI \citep{semb03}.  These absorption lines are at
zero redshift, so the gas could either be a Galactic halo or a Local Group
medium.  If it is a Galactic halo (typically 50 kpc or smaller), the mass
involved is less than the known mass of cold gas in galaxies, and this is
argued for by \citet{wang05} and \citet{breg07}.  However, if the gas fills
the Local Group, then the hot gas mass can dominate the total baryon
content \citep{nica02,nica03,rasm03}, as expected from the missing baryons. 
Since there is also diffuse X-ray emission that is associated with this gas
\citep{mccam02}, one can hope to use the shadowing technique, provided
that there are absorbing clouds at distances of at least tens of kpc from
the Milky Way.  There are such clouds, as the Magellanic Stream has
some dense knots of \ion{H}{1} and it lies at about 50--60 kpc.
Another class of potential targets are the compact high velocity clouds, 
which \citet{blitz99} suggested to be orbiting through the Local Group 
at distances of $\sim10^{2.5}$ kpc.  However, new evidence suggests that these
are associated with the Galactic halo and lie within 50 kpc \citep{malo03,putm03b, west07}.
We have obtained observations toward these clouds suitable for shadowing
studies and report on them below.

\section{Observations}

There are five different observations discussed below of two edge-on
galaxies and two \ion{H}{1} clouds in the Local Group.  As some observations
were obtained with {\it Chandra\/} while others obtained with {\it XMM\/}, we
will discuss the results as a function of the instrument used.

\subsection{{\em Chandra\/} Data and Data Reduction}

The {\it Chandra\/} observations that we analyze in this work were obtained
with the {\it ACIS\/} instrument with the objects centered either near the
aimpoint of the S3 chip or near the center of the S3 chip.  Basic
observational information on the individual pointings are given in Table 1.

For the data reduction, which has become fairly standard, we used CIAO
3.2.2 along with CalDB 3.1.0.  In each set of data, we identified the bad
pixels and pixels affected by cosmic ray afterglows and excluded them
from the data set; only grades 0, 2, 4, and 6 were used.  Periods of high
background were excluded, with the threshold being 40\% above the
modal value of the background.  Exposure maps were produced for the S3 chip, on
which our analysis was focused, and the images were either flattened or the
exposure map was used separately to correct the extracted counts.  For
the identification of point sources, we used both {\it wavdetect\/} and direct
identification by eye, which led to nearly identical results.  The
observations of the Magellanic Stream Cloud were obtained in FAINT
mode while the observation of NGC 891 and the Compact High Velocity
Cloud were obtained in VFAINT mode.

\subsubsection{Magellanic Stream Cloud MS30.7-81.4-118}

\begin{figure}
\plotone{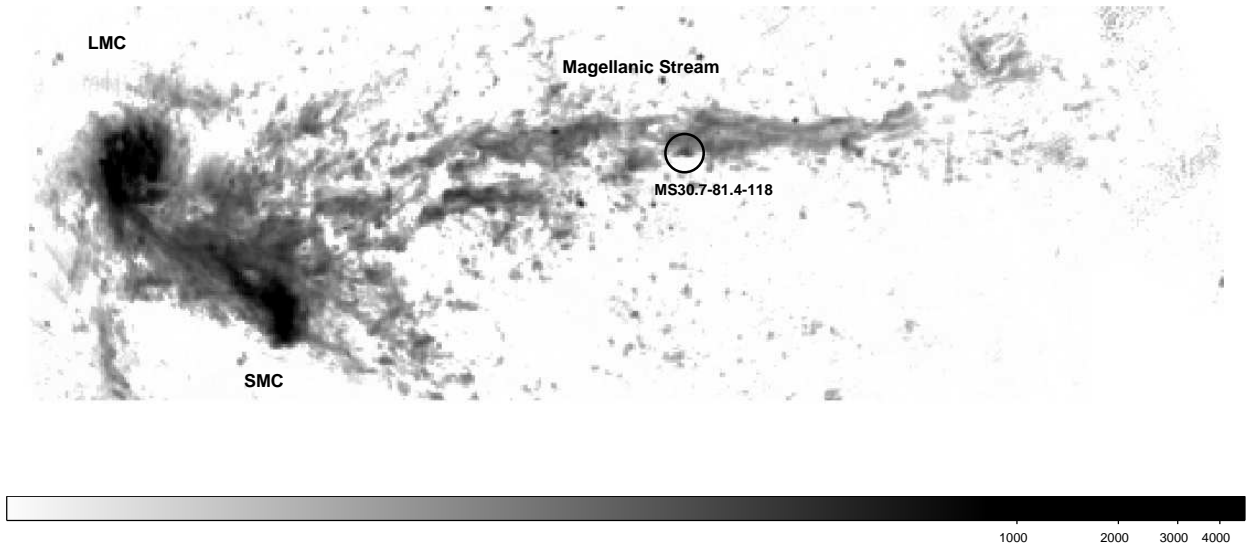}
\caption{The 21 cm total intensity map of the LMC, SMC, and the Magellanic
Stream \citep{putm03a}, with the location of cloud MS30.7-81.4-118 marked with 
a circle of radius 2$\arcdeg$.  The intensity values are on a logarithmic scale 
with black corresponding to $N_{HI} > 6{\times}10^{20} cm^{-2}$, and the faintest 
levels corresponding to $\sim 2{\times}10^{18} cm^{-2}$.  \label{MStream}}
\end{figure}

\begin{figure}
\plotone{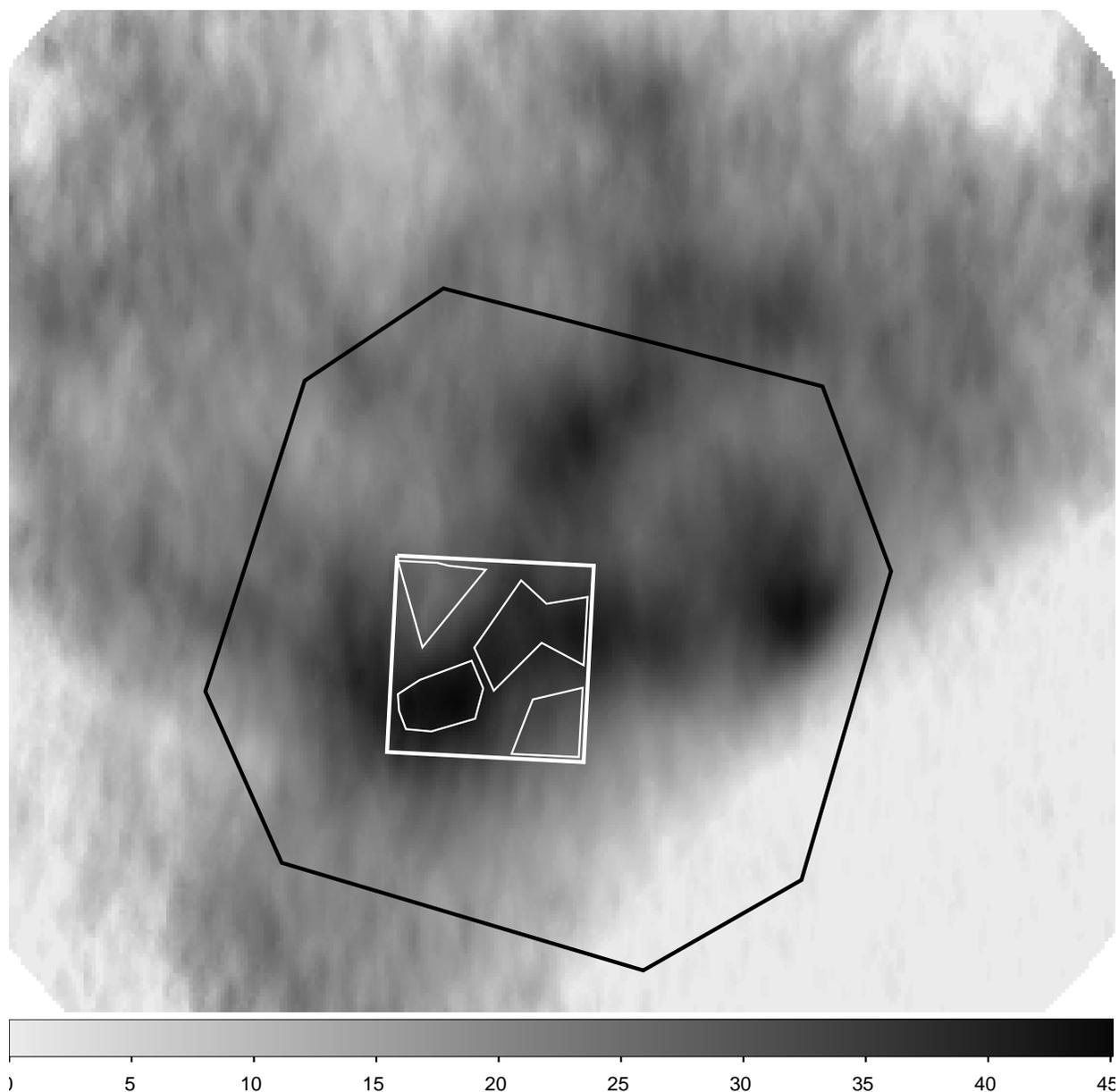}
\caption{The 21 cm total intensity map of MS30.7-81.4-118 with the
{\it Chandra\/} ACIS-S3 field of view (8$\arcmin$ square) and the approximate
{\it XMM-Newton\/} EPIC field of view (larger polygon) superimposed.  The four
polygons within the ACIS-S3 field show the high and low column density regions
used for the extraction of the X-ray counts.  The scale at the bottom is in
units of 10$^{19}$ cm$^{-2}$.  In the region of the X-ray observations, the peak
\ion{H}{1} column density is 4$\times$10$^{20}$ cm$^{-2}$,
about three times higher than the foreground Galactic column density of
1.5$\times$10$^{20}$ cm$^{-2}$.  \label{MS30fovs}}
\end{figure}

One of the best targets is a neutral gas cloud at a distance near the edge
of what might be considered the Galactic Halo, about 50--70 kpc.  At this
distance is a set of clouds, the Magellanic Stream \citep{gard96}, 
which passes over the South Galactic Pole. Recently, a deep
survey was obtained with the Parkes 21 cm multibeam system \citep{putm03a},
revealing a great deal of structure in the Stream at 15$\arcmin$ 
resolution (Figure \ref{MStream}). 
Higher resolution (1$\arcmin$) radio synthesis observations were obtained by \citet{bruens98} of the highest column density regions and one region
shows structure with \ion{H}{1} columns exceeding 4$\times$10$^{20}$ cm$^{-2}$
(Figure \ref{MS30fovs}), significantly greater than the Galactic \ion{H}{1} column of
1.5$\times$10$^{20}$ cm$^{-2}$.

Two partly-overlapping images were obtained, on the \ion{H}{1} maximum, and
away from the \ion{H}{1} maximum.  The exposure on the \ion{H}{1} maximum was
one of the rare times when there were no flares and the background was
relatively low, so nearly the entire exposure was used and this was the most
useful of the two exposures.  Potential shadowing by the Magellanic
Stream cloud should have the most pronounced effect at energies below 1
keV, so we produced a flattened 0.4--1.0 keV image with point sources
removed (they accounted for 20\% of the total counts in the image).  In
this image, there is an apparent extended symmetrical brightening at
00:13:32.6, -27:11:07 (RA, Dec, J2000), which may be a background
group or cluster; it appears in both the soft and hard (1--6 keV) bands.

To search for shadows, four regions were chosen, from the highest to the
lowest \ion{H}{1} regions.  Using the flattened hard image, all four regions
were consistent with a constant value for the surface brightness, being within
1.3$\sigma$ of the mean.  In the 0.4--1.0 keV image, there is some modest
variation, with the high \ion{H}{1} column regions being brighter than the low
\ion{H}{1} regions, the opposite of the shadowing signature.  When we compare
the two highest \ion{H}{1} regions to the two lowest \ion{H}{1} regions, the
surface brightness values are 1.29 $\pm$ 0.04 cts ksec$^{-1}$ arcmin$^{-2}$
and 1.20 $\pm$ 0.04 cts ksec$^{-1}$ arcmin$^{-2}$, a difference of
0.093 $\pm$ 0.058 cts ksec$^{-1}$ arcmin$^{-2}$.  This difference
corresponds to a modest brightening of 1.6$\sigma$, rather than the dimming that
a shadow would produce.  

There will be fluctuations in the X-ray sky due to variation in the Galactic
X-ray background, which we can assess in this part of the sky by using
our off-cloud data.   Even with a longer observation, it may not be
possible to improve the limit on the shadow due to the fluctuations in the
soft Galactic X-ray background (this region has a fairly smooth
background, as seen from the {\it ROSAT\/} data \citep{snow97}.  We analyzed the off-cloud
images, which after screening high background regions (more than 50\%
higher than the modal value), led to a useful observing time of 33.49 ksec. 
After flattening the image, we divided it into nine regions about 300 pixel
square, about 15\% smaller than most of the regions used for the on-cloud 
study.  The errors per bin were about 6\% and the 0.4--1.0 keV
background in the field is consistent with a constant value, as the
variation is $\pm 1.4{\sigma}$.

Next, we need to separate the amount of flux from the
instrument and from the X-ray background.  The instrumental background for
the ACIS S3 chip was fairly stable over the past five years (since 2000.5)
and lower than at launch.  The total chip background for the observations
is expected to be fairly modest (and similar) for the observation dates of
our targets (7.5--9 cts s$^{-1}$ chip$^{-1}$; see Figure 6.24 in MPOG v8).
Using the standard grades, avoiding flares, and for the 0.4--1.0 keV band, the
instrumental count rate is expected to be about 4.2$\times$10$^{-8}$  cts
pixel$^{-1}$ for 0.5$\arcsec $ pixels, or 0.61 cts ksec$^{-1}$ arcmin$^{-2}$.  
This corresponds to about half of the counts registered in the above
measurements.

Another component that needs to be subtracted is the unresolved point source
contribution from AGNs.
For a 50 ksec Chandra ACIS-S observation, where the limiting point source is 
defined by five photons, the unresolved point source contribution to the diffuse
emission is about 15\% of the point source component of the background in the
0.5--2.0 keV band \citep{more03}.  
When converted to the 0.4--1.0 keV band, the unresolved
point source contribution is 0.037 cts ksec$^{-1}$ arcmin$^{-2}$.
This unresolved point source contribution is less than 10\% of the diffuse 
background for all Chandra observations and closer to 5\% for most of the
Chandra observations.  We remove this small component, correctly scaled to
the observing time, in our analysis.

We remove these various components to obtain the diffuse emission from gas 
in the shadow and non-shadow region.  With minor differences, this procedure
is followed for the other observations.
There is one final consideration, which is the removal of the emission 
from the Local Hot Bubble (LHB).  Once this is removed (discussed below), we 
have measurements of true diffuse emission from above the disk.

\subsubsection{CHVC 125+41$-$207}

Most of the high velocity clouds are resolved easily 
with single dish radio telescopes, but there are a number that have
significant columns yet have small angular sizes.  These are the compact high
velocity clouds (CHVC).  \citet{braun99,braun00} mapped several CHVCs and a few of them
have column densities of up to 4x10$^{20}$ cm$^{-2}$, with sizes of 10--50
arcmin$^2$.  These clouds are ideal targets for X-ray shadowing since they are
mapped at good S/N in \ion{H}{1} and have high density regions that fill a significant
fraction of a single ACIS chip. 

\begin{figure}
\plotone{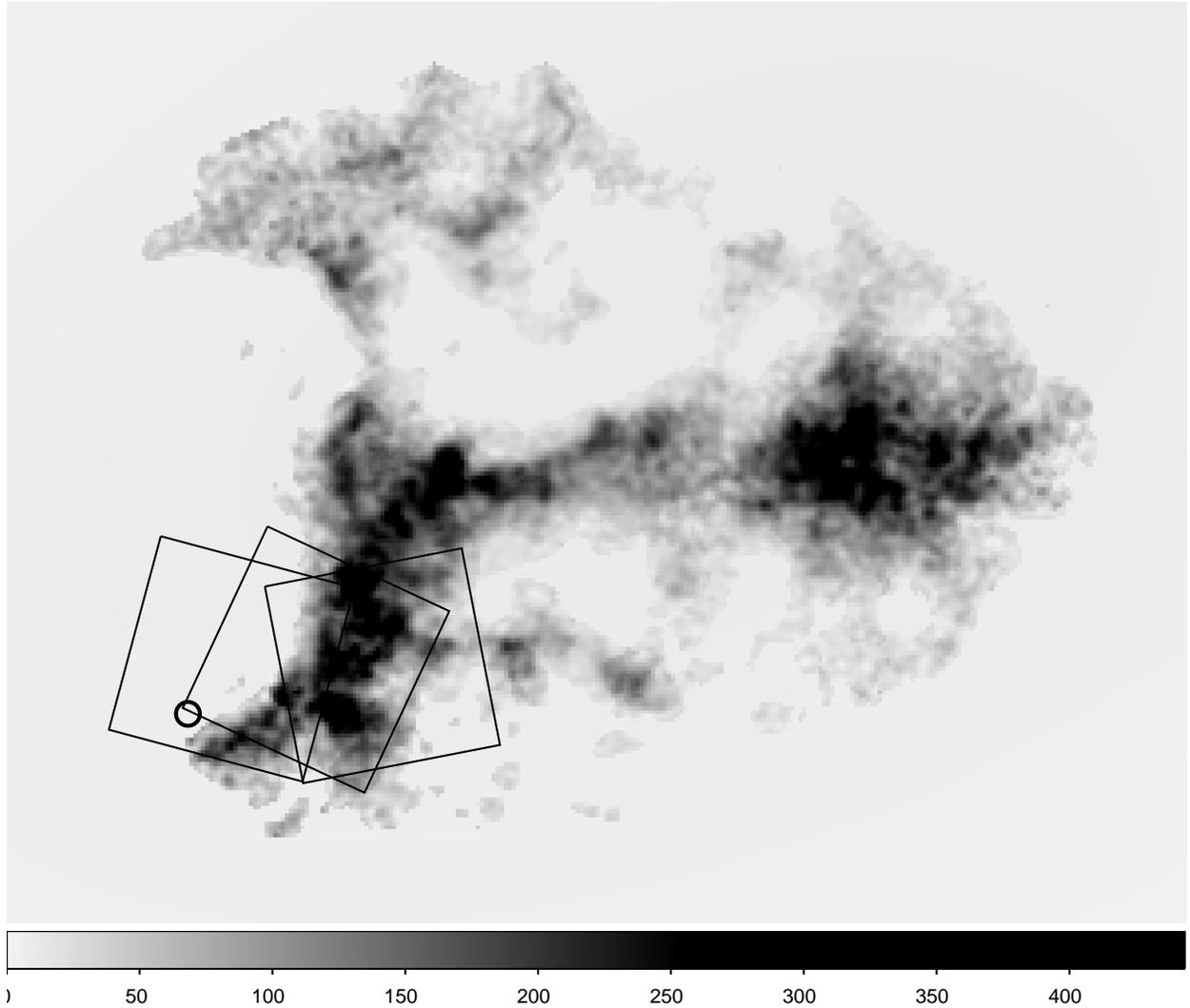}
\caption{The 21 cm total intensity map of CHVC 125+41$-$207 with the
{\it Chandra\/} ACIS-S3 field of view (8$\arcmin$ square) superimposed for
ObsID 2253, 2484, and 2486 (left to right); the scale at the bottom is in units
of 10$^{18}$ cm$^{-2}$.
In this interferometric \ion{H}{1} column density map \citep{braun00}, the
peak cloud intensity within the ACIS-S3 field is 5$\times$10$^{20}$ cm$^{-2}$,
while the regions adjacent to the cloud have a mean value of about
3$\times$10$^{20}$ cm$^{-2}$.  The circle marks the location and approximate
size of the diffuse enhancement in the X-ray emission.  \label{CHVCfovs}}
\end{figure}

\begin{figure}
\plotone{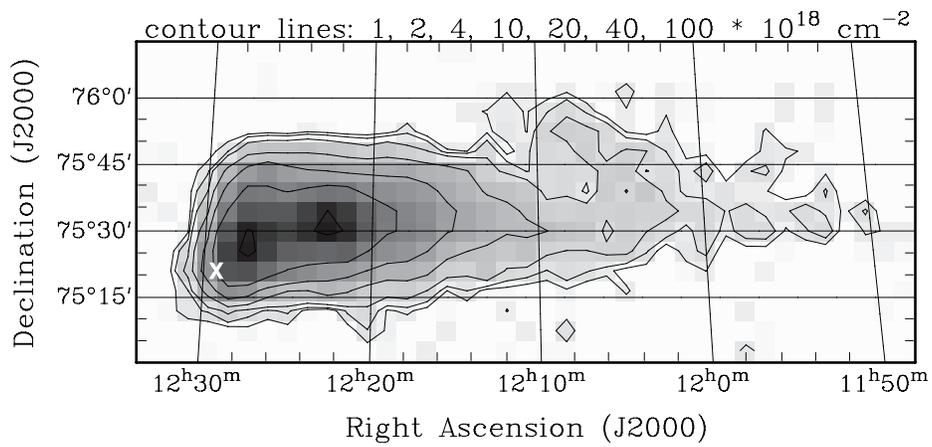}
\caption{The single-dish 21 cm total intensity map of CHVC 125+41$-$207  
shows that the cloud has a head-tail configuration \citep{bruen01}.  
The highest column density regions are observed in the X-ray band (see Figure 3).
The location of the diffuse X-ray enhancement is shown with a white X (also shown above).
 \label{CHVC_single_dish}}
\end{figure}

In selecting a CHVC as a target, we needed a cloud whose column
density is comparable to that of adjacent regions in the Galaxy and whose
solid angle covers a significant fraction of the ACIS S3 chip, which is the
most advantageous for this program.  Of the clouds that have been
mapped with radio synthesis arrays in the 21 cm \ion{H}{1} line \citep{braun00}, 
the cloud CHVC 125+41-207 is the most suitable (see Figure \ref{CHVCfovs}).  
It is at {\it l, b\/} = 125\degr , 41\degr\ (and velocity -207 km s$^{-1}$), 
with the highest column density part confined to an
elongated structure of dimensions $6\arcmin \times 17\arcmin$.  At lower column
density levels, single-dish data show that it has a head-tail configuration (Figure
\ref {CHVC_single_dish}) and is about 0.8\degr\ by 2.5\degr\ \citep{bruen01}.
The synthesis \ion{H}{1} map, which has 28$\arcsec$ resolution, shows that the highest
column density part of the cloud is about 5$\times$10$^{20}$ cm$^{-2}$, which is
larger than that of smooth regions adjacent to the cloud (3$\times$10$^{20}$ cm$^{-2}$),
which are due to the Galactic disk. Aside from the usual point sources, there is a
previously unknown diffuse feature that occurs just to the east of the \ion{H}{1} 
cloud with an RA and DEC (J2000) of about 12:29:38.7, 75:21:00, 
and with a radius of about 30\arcsec .
This falls in the background subtraction region of the image and was excluded
when quantifying the shadow.  There is no obvious optical counterpart on the 
Digital Sky Survey images, so it is most likely either a distant cluster or it
is caused by the interaction of the CHVC with its surroundings.

This cloud was observed three times at different positions and roll angles. 
In all cases, we identified and excluded point sources from the
analysis and defined on-cloud and off-cloud region.
In the first observation (Obsid 2484), there were several long periods of
high background, so the final useful data was 10.1 ksec.  In the second
observation (2486), the instrumental background was well-behaved and
we were able to use most of the data, 21.2 ksec.  Conditions were most
favorable in the last observation (2253), which is twice as long as the
preceding observation and the orientation of the chip is ideal so that about
half contains high-column \ion{H}{1} and half has the lowest \ion{H}{1} column
of any chip position; the useful time was 45.3 ksec.  The three observations
yield a useful integration time of 76.5 ksec, so we differenced the regions of
higher \ion{H}{1} column (5$\times$10$^{20}$ cm$^{-2}$) and lower \ion{H}{1}
column (3$\times$10$^{20}$ cm$^{-2}$). 
The mean surface brightness in the higher \ion{H}{1} regions is 
1.20 cts ksec$^{-1}$ arcmin$^{-2}$, 
which was 4.9\% brighter than the surface brightness of the
lower column density region.  As the uncertainty in the difference is
3.8\%, the brightening is only a 1.3$\sigma$ effect, although in the opposite sense of the
shadow that we were searching for.  

\subsubsection{NGC 891}

The more recent and longer observation of NGC 891 had only a few
periods of high background so that 108.5 ksec were used out of the total
exposure time of 120.4 ksec.  Not only is this useful time a factor of
three greater than the 36 ksec used in the earlier exposure, but the
instrumental background appears to be lower.  In addition, two strong
sources in NGC 891 are seen in both exposures (one source is SN
1986J), so a very high relative accuracy is achieved and we can be certain
that the regions used to analyze the earlier observation are sampling the
same J2000 coordinate regions.

\begin{figure}
\plotone{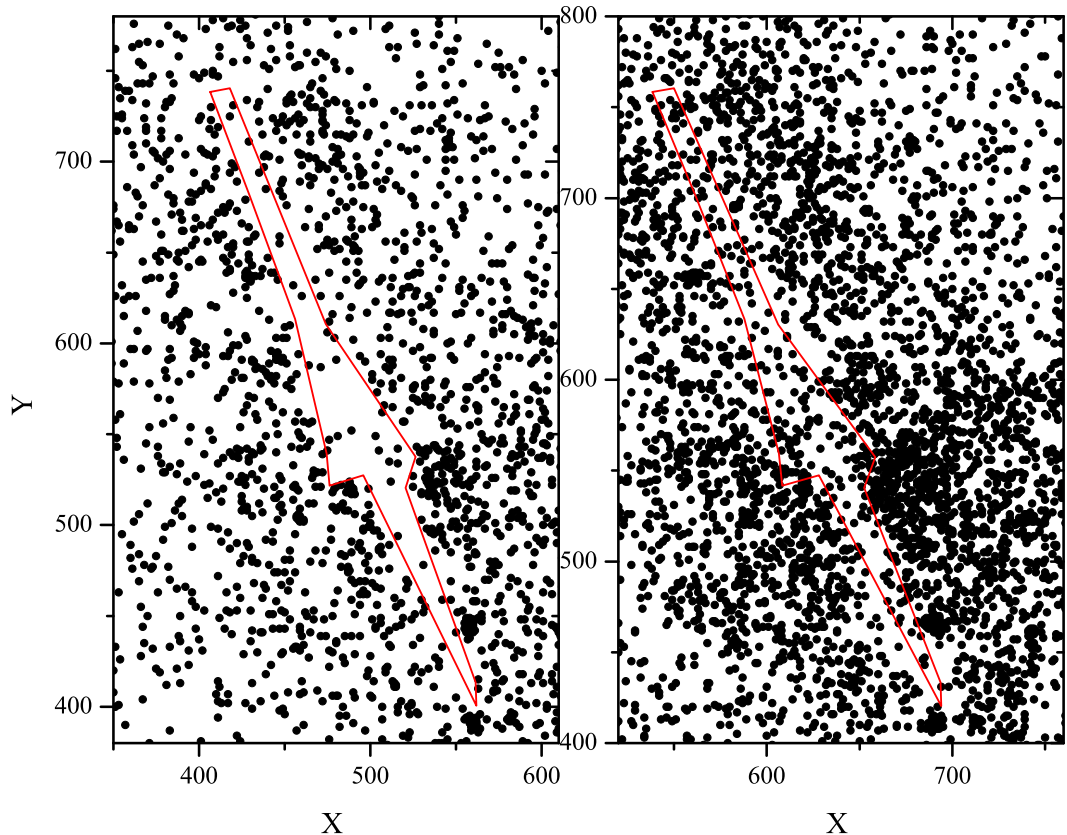}
\caption{Two {\it Chandra\/} images of NGC 891 in the energy band 0.4--1.0 keV,
where the region of high optical extinction is delineated by the polygon.  Each
pixel is 0.5$\arcsec$ and the field is 3.3$\arcmin$ high.  Each dot is a single
photon.  The X-ray shadow seen in the first observation (left, \citealt{breg02})
is not confirmed with more recent data (right). \label{n891shadows}
}
\end{figure}

The shadow region was defined in \citet{breg02} by the dark extinction
region that obscures nearly all of the optical light from parts of the disk. 
All point sources were excluded, an important consideration because
there are several point sources detected from the disk, most of which are
probably X-ray binaries.  The extinction region, with the point
sources masked out, constitutes 1378 arcsec$^2$ (see Fig. \ref{n891shadows}).
For the background region, we needed to obtain a surface brightness that would
not be contaminated significantly by the extended emission from NGC 891, so
these regions are in the outer part of the S3 chip, perpendicular to the
disk of the galaxy.  The exposure corrected background on the SE side is
1.22 $\pm$ 0.04 cts ksec$^{-1}$ arcmin$^{-2}$ and on the NW side it is
1.39 $\pm$ 0.04 cts ksec$^{-1}$ arcmin$^{-2}$,
so there may be a slight gradient in the Galactic background; the
average of the two backgrounds is 1.31 $\pm$ 0.08 cts ksec$^{-1}$ arcmin$^{-2}$.
In comparison, the surface brightness in the extinction region is 1.80 $\pm$
0.24 cts ksec$^{-1}$ arcmin$^{-2}$, a difference of 0.49 $\pm$ 0.25 
cts ksec$^{-1}$ arcmin$^{-2}$, which is 2$\sigma$
{\it above\/} the background level, whereas the earlier observation had found it
to be about 2$\sigma$ {\it below\/} the background level.  This slight excess is
also present when we subdivide the energy bands into 0.4--0.7 keV and 0.7--1.1
keV (about a 1.4$\sigma$ excess in both bands).  Thus, the shadow previously
seen in NGC 891 is not confirmed and we conclude that the earlier
observation was a statistical fluke.

\subsection{{\em XMM\/} Data and Data Reduction}

\subsubsection{NGC 5907}

\begin{figure}
\plotone{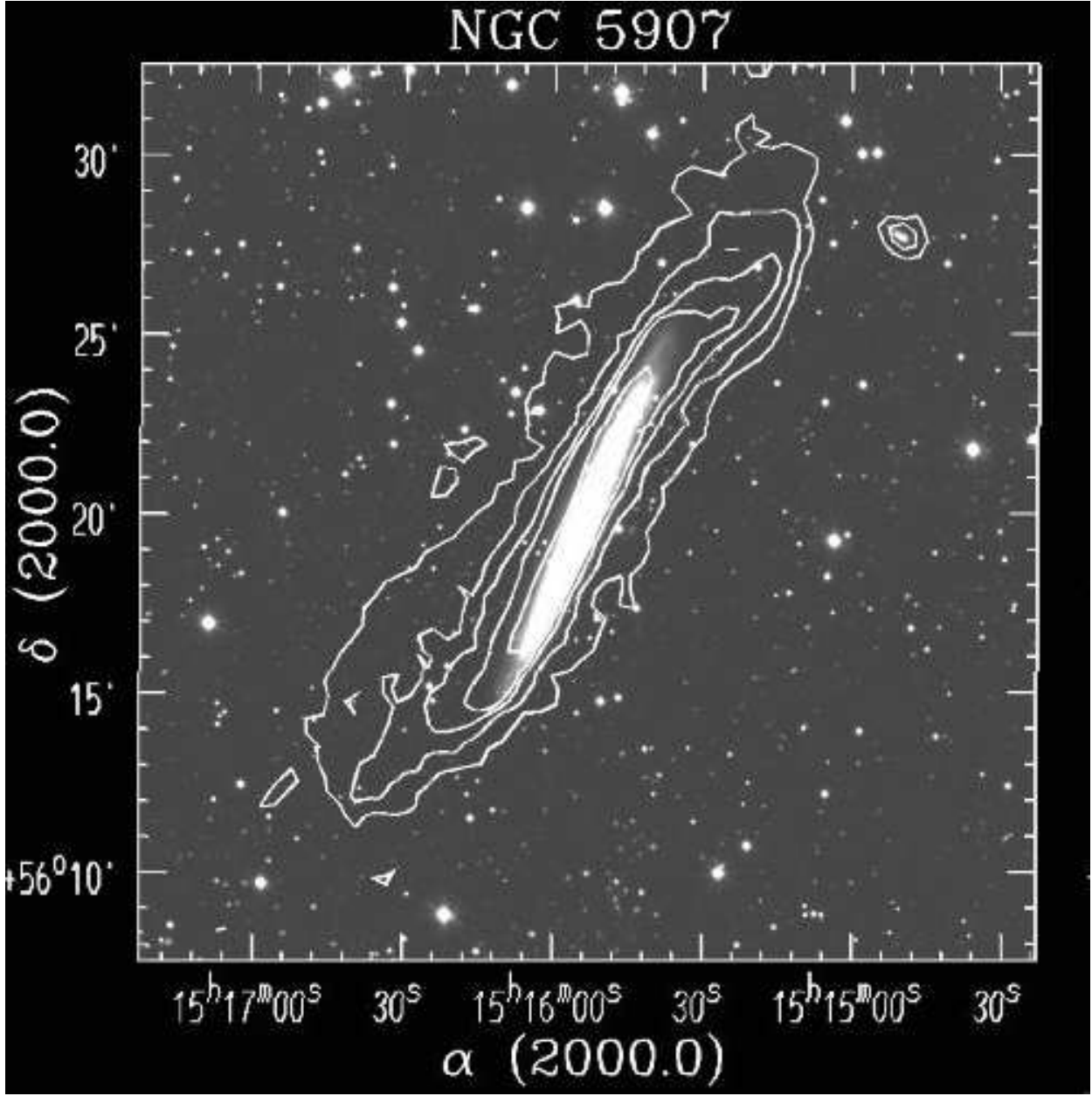}
\caption{The \ion{H}{1} contours are superimposed on the optical image of the
edge-on galaxy NGC 5907 \citep{sanc76,shang}.  The region where the \ion{H}{1}
extends beyond the optical galaxy is where we have sought to search for a shadow
in the diffuse X-ray background that is produced from absorption by the cold
gas.  \label{n5907optHI}
}
\end{figure}

We have examined the \ion{H}{1} maps for all edge-on galaxies as well as the gas
torn away in interacting galaxies and we found that the best combination
of optical depth and solid angle is given by the \ion{H}{1} in NGC 5907.  The
gaseous disk of NGC 5907 not only extends beyond the stellar disk, it has
a modest warp where the gas turns away from the smaller stellar disk 
(Figure \ref{n5907optHI}).  The column density of this extended gas rises 
above $1 \times 10^{21}$ cm$^{-2}$, which causes significant absorption 
in the 0.4--1.0 keV region. 

The observations of NGC 5907 were taken February 20 and 28, 2003
(observations 2 and 1, respectively, with total exposure times of 52.74
ksec and 52.19 ksec). The data were extracted using the Science Analysis
Software (SAS) version 7.0.0 of the {\em XMM-Newton\/} group. Unfortunately,
the observations were contaminated by strong flaring. We excluded the time
intervals of the flares and in addition applied the standard three--sigma
clipping to the data based on the 12--14 keV light curves. We checked the
resulting good time intervals in the lower energy band (0.4--7 keV). No
additional time intervals had to be excluded. Due to the strong flaring, only
12.5 ksec of exposure time were usable for NGC\,5907 (about 12\% of the total
exposure time). For our analysis, we used only the PN data because of the higher
sensitivity of that CCD.

The data were binned by 50 pixels during SAS processing. The lowest energy range
(0.2--0.4 keV) is dominated by an instrumental background, whereas the
background in the highest energy range (1.0--10.0 keV) is expected to consist
mostly of unresolved AGNs. We therefore extracted a 0.4--1.0 keV energy band
image for our shadowing experiment. In addition, a 0.5--2.0 keV and a large
energy band image (0.2--10.0 keV) were also created to aid in our analysis.

In order to identify point sources we created two smoothed images for either
observation by applying Gaussian filters of 5 and 25 binned pixels,
respectively, to the 0.2--10.0 keV images. The difference images between these
two smoothed images revealed point sources and small scale structures from which
we created a bad pixel mask for either observation. Remaining possible point
sources were marked by hand and included in the bad pixel mask. Dead columns
were also identified and masked. CCD boundaries and hot pixels were derived from
the exposure maps and included in the bad pixel masks as well, which were then
applied to the filtered images.

In this shadowing experiment, a flat background is crucial to obtain
reliable measurements. While part of the X-ray background is caused by scattered
high energy particles creating a smooth background $B_{\rm flat}$, high energy
particles entering the aperture and following the optical path are subject to
the vignetting effects of the mirrors creating a focused background
$B_{\rm foc}$. We attempted to disentangle these two components in each energy
band by measuring the background counts $C_{\rm i}$ (i = 1--15) in 15 circular
aperture (each about 100\arcsec\ in diameter) spread across the NGC\,5907
observation away from a possible galaxy X-ray halo. We also obtained the
exposure times $t_{\rm i}$ for these regions using the exposure map. The flat
background then could be derived for each set using the least $\chi^2$ method on
$$ C_{\rm i} = B_{\rm foc}\cdot t_{\rm i} + B_{\rm flat}. $$

\begin{figure}
\plotone{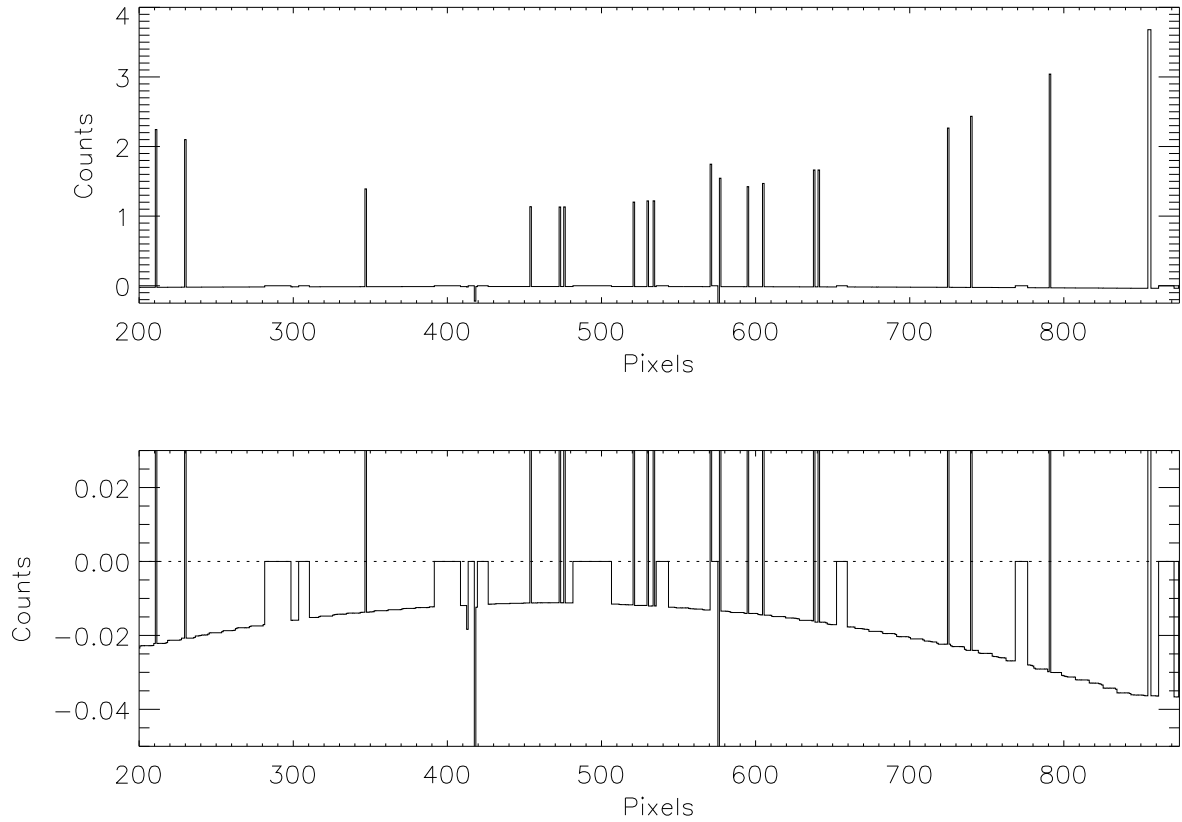}
\caption{\label{flat}Horizontal cross-section of the flattened 0.4--1.0 keV
image of NGC\,5907. Both panels show row 510 near the center of the image; the
top panel covers the full range of the positive pixel values, wherease the
bottom panel focusses on the pixels that did not receive any photon events and
now posess negative values after the flattening. The dotted line represents zero
counts for easier reference. Bad pixels are masked and set to zero after the
flattening. \label{ngc5907slice}}
\end{figure}

We subtracted the $B_{\rm flat}$ component, divided by the exposure map and
normalized the image by multiplying it with the actual exposure time of 12.5
ksec. The resulting flattened image displays an apparent inverse vignetting
pattern (e.g., positive pixels values increase towards the outer parts of the
field of view as shown in Fig. \ref{ngc5907slice}). To understand this pattern, one has
to consider the fact that most pixels of the X-ray background in our data did
not receive any photon events even after the binning. The vignetting pattern in
the data is therefore not represented by an increase in pixel values toward the
optical axis, but by an increase in photon event density over a larger area. In
the flattened image, the former non-zero pixel values therefore increase toward
the edges of the CCD, whereas the former zero-value pixels are decreasing to
more negative values at the edge of the field of view. The larger density of
negative pixels thus compensates the increase in positive pixel values toward
the edges of the flattened image.

For extended sources filling the field of view (i.e., MS30.7), this
method of deriving the flat background component is not possible. The basic vignetting pattern cannot be removed from observations of
extended sources (this is a more significant problem than in the {\it Chandra\/}
observations that used only a single chip with a smaller field of view). We
therefore tried a second method using the closed filter wheel (CFW) data of
\citet{marty} to derive the flat background component. For this CFW method, we
chose measurement regions along contours of constant effective exposure time to
exclude the vignetting effects from our measurements. The instrumental
background shows spatial inhomogeneities at certain energies due to the
electronics board behind the CCD \citep[e.g.,][]{frey}. Most of these
fluorescence lines, however, are at higher energies and thus should not affect
the 0.4--1.0 keV band. We measured the instrumental background in the same
regions of the CCD as for NGC\,5907 and subtracted those count rates from our
galaxy measurements accordingly. For better comparison, we chose the same
measurement regions for both methods.

\begin{figure}
\plotone{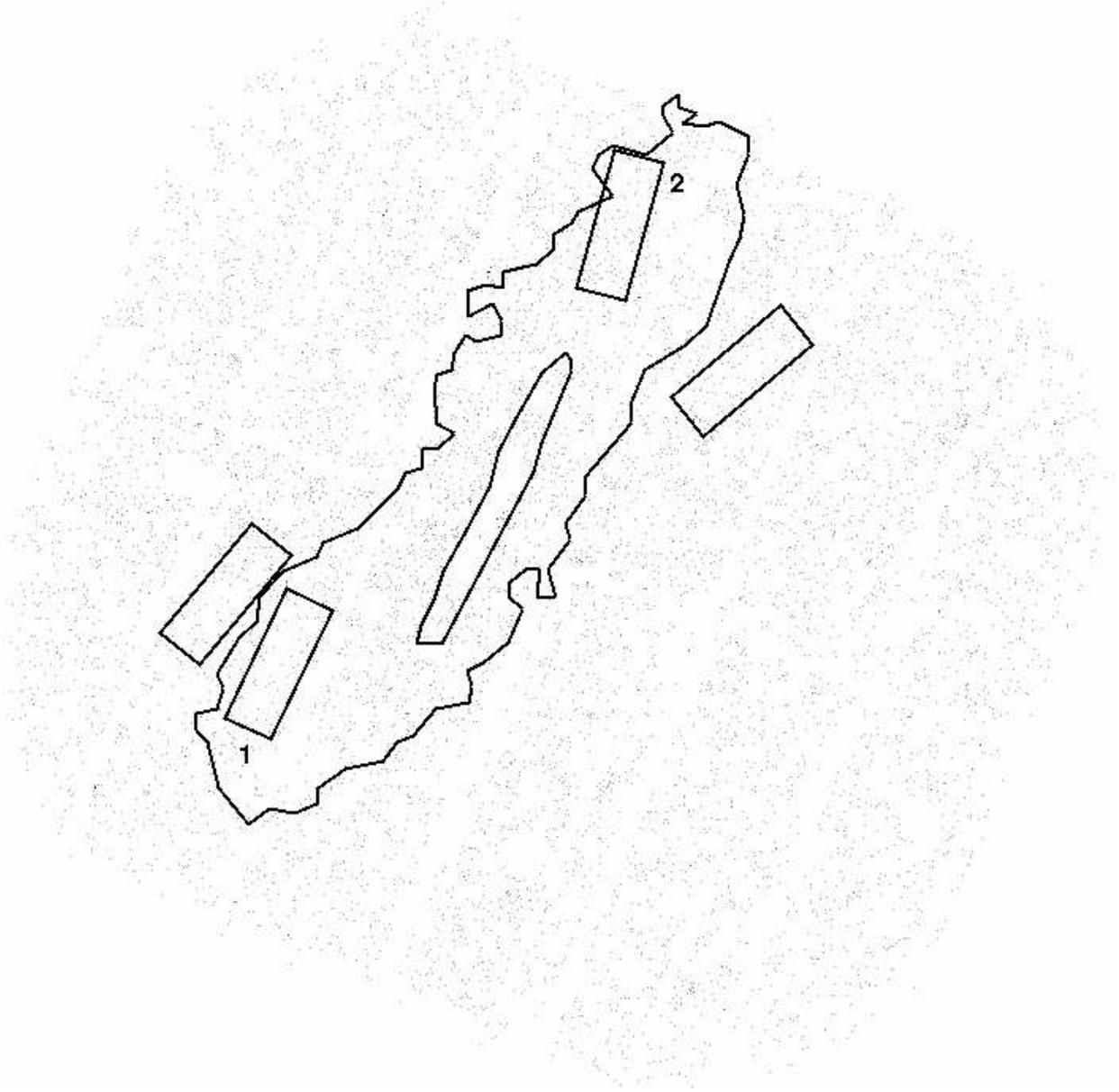}
\caption{\label{n59}{\em XMM\/}-PN 0.4--1.0 keV image of NGC\,5907. A bad pixel
mask has been applied setting corresponding pixel values to zero. The shown
\ion{H}{1} contours \citep[0.69 and 75 mJy~beam$^{-1}$,][]{shang} represent the
optical disk of the galaxy and the \ion{H}{1} halo. The two pairs of rectangular
regions used for the X-ray background measurements are shown as well. \label{ngc5907reg}}
\end{figure}

We used the \ion{H}{1} contours of \citet[][our Fig. \ref{n5907optHI}]{shang} to
determine
the area that could absorb X-ray background photons. The area of the galactic
disk has the highest absorbing column, but it is also more likely to be
contaminated by unresolved X-ray point sources. We therefore chose regions in
the \ion{H}{1} halo as areas of possible shadowing and regions outside the halo
as X-ray background. We tried to make the measurement regions as large as
possible while minimizing the number of masked pixels in each region (as those
may have different effects on the count rate due to the vignetting). Figure
\ref{ngc5907reg} shows the two region pairs, one in the northern and one in the
southern part of the field of view. The number of actual pixels in each region
was obtained from the bad pixel mask to correct the measured photon counts. The
measurements of both regions are listed in Table \ref{n59meas} for both methods.

Regions 1 and 2 yielded on-minus-off-galaxy measurements of
$-0.45\pm0.21$ and $+0.29\pm0.23$ cts ksec$^{-1}$
arcmin$^{-2}$, respectively, for the $\chi^2$ method and
$-0.60\pm0.34$ and $+0.34\pm0.39$ cts ksec$^{-1}$
arcmin$^{-2}$ for the CFW method.  This indicates a deficit in the sense of a shadow at the 2$\sigma$ level in the southeastern part of the galaxy halo, whereas the
northwestern halo shows no shadow. Assuming a smooth X-ray background in the
northwestern halo (and thus averaging the on- and off-measurements), we derive a
3$\sigma$ upper limit of 21--26\% for a shadow in this region depending on the
method used for the flattening.

The detection limit for point sources in our combined observations of NGC\,5907
in the 0.5--2.0 keV band is $3.6\times10^{-15}$ ergs s$^{-1}$ cm$^{-2}$. Based
on the source distribution of \citet{more03} in the 0.5--2.0 keV band and our
detection limit, we derive a contribution of unresolved AGNs and galaxies of
about 33\% in our observations. This corresponds to a background
flux\footnote{Conversion between count rates and fluxes was performed in
WebPIMMS at http://heasarc.gsfc.nasa.gov/Tools/w3pimms.html using a power law
index of 1.4.} for unresolved sources of about $0.25$ cts ksec$^{-1}$
arcmin$^{-2}$ in the 0.4--1.0 keV band. When this is removed from the potential shadow, it becomes a statistically insignificant 1$\sigma$ fluctuation.
\subsubsection{MS30.7-81.4-118}

\begin{figure}
\plotone{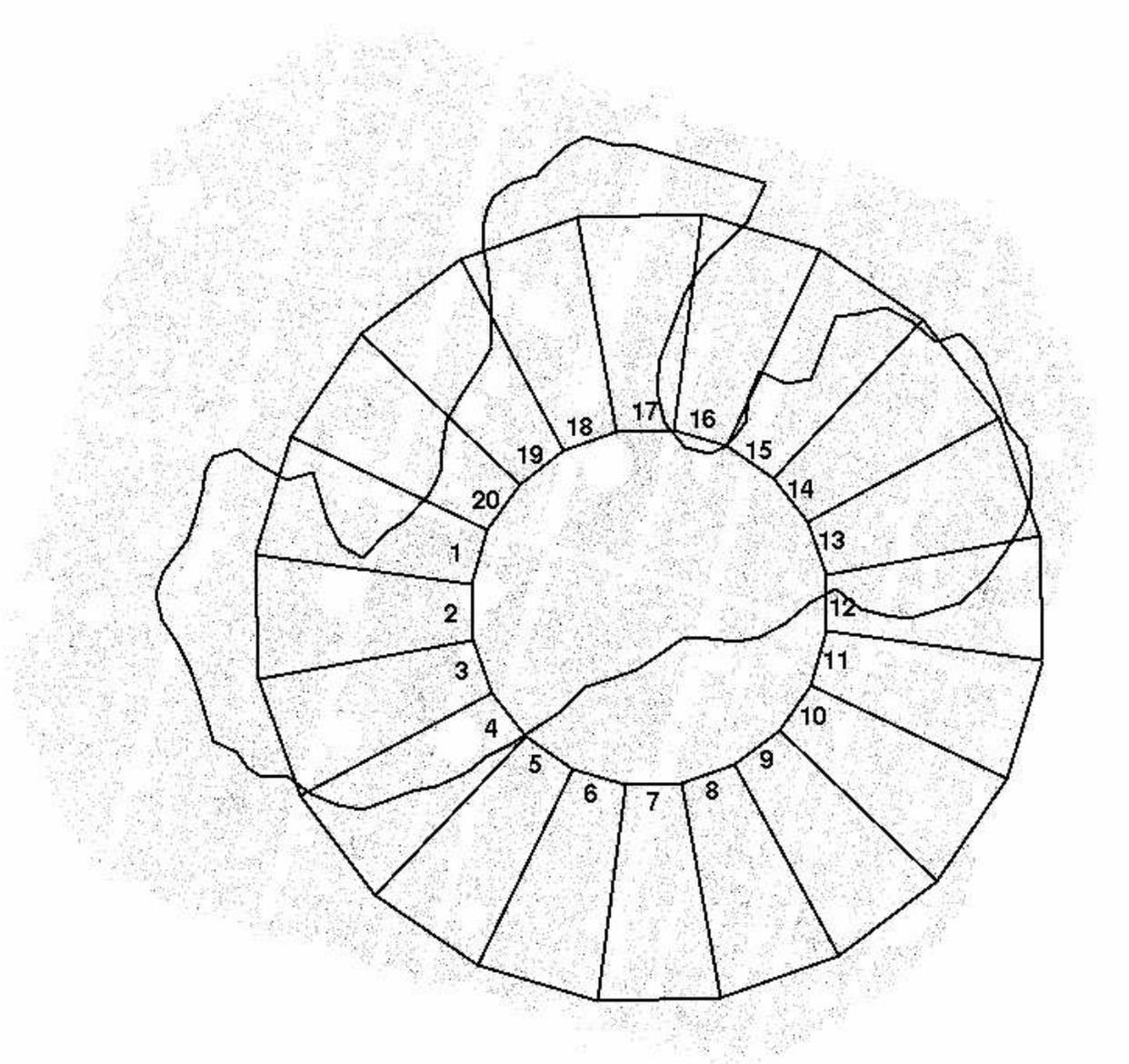}
\caption{\label{msc}{\em XMM\/}-PN 0.4--1.0 keV image of the Magellanic Stream
Cloud MS30.7-81.4-118. A bad pixel mask has been applied setting corresponding
pixel values to zero. Only one \ion{H}{1} contour
\citep[$N({\rm HI})=3\times10^{20}~{\rm cm}^{-2}$,][]{bruen05}
is shown. The \ion{H}{1} column density drops sharply on the southwest side of
the cloud, but decreases only slowly toward the northeast. The twenty
measurement regions are labeled and follow the exposure map contours. \label{ms30.7creg}}
\end{figure}

\begin{figure}
\plotone{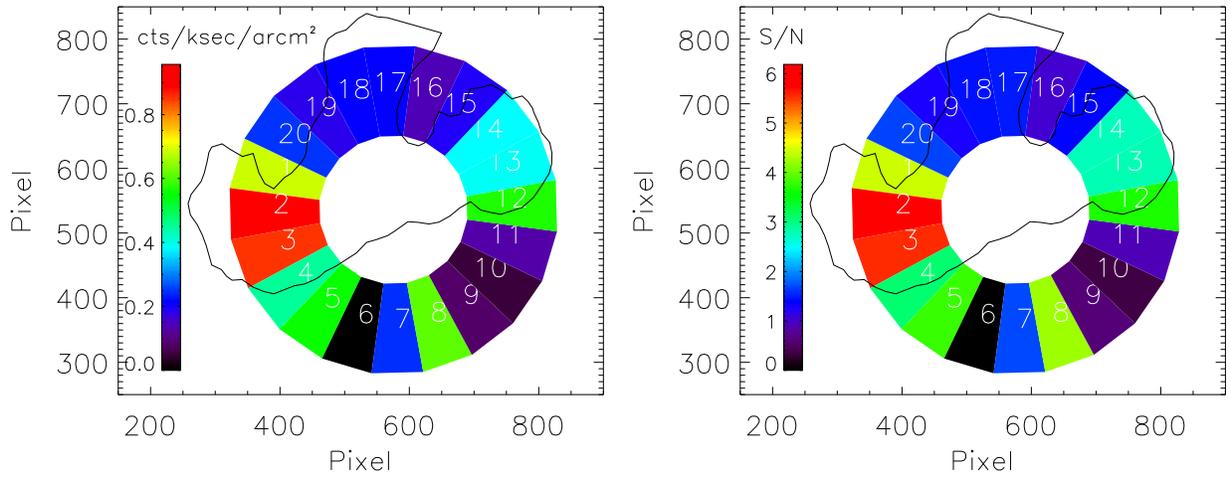}
\caption{\label{mscmap}{\em XMM\/}-PN measurements of the Magellanic Stream
Cloud MS30.7-81.4-118. The left panel shows the X-ray fluxes in cts ksec$^{-1}$
arcmin$^{-2}$ relative to region 6 (whose flux is thus zero). The right panel
shows the resulting signal-to-noise. The black outline in either panel
represents the \ion{H}{1} contour $N({\rm HI})=3\times10^{20}~{\rm cm}^{-2}$
\citep{bruen05}. \label{msc30.7map}}
\end{figure}

The observation of the Magellanic Stream cloud MS30.7 was taken January 3, 2004.
The data extraction and calibration was the same as for NGC\,5907 described
above. The light curves of MS30.7 showed little flaring, resulting in 37.0 ksec
of good time intervals (71\% of the total exposure time). Due to the extent of 
the \ion{H}{1} cloud on the {\em XMM\/} field of view, we did not try to
derive the smooth background component using the least $\chi^2$ method described
above. We again chose measurement regions that followed constant exposure time
contours of the exposure map to exclude vignetting effects and applied the CFW
method to measure the background. Our first measurements yielded an X-ray excess
on part of the cloud. We therefore mapped the X-ray emission along the segments
of a circle centered on the exposure map peak (Fig. \ref{ms30.7creg}). Table
\ref{mscmeas} lists the individual fluxes of all 20 regions and the flux
differences relative to region 6, which has the lowest flux of all the regions.
Fig. \ref{msc30.7map} displays these flux differences to emphasize the X-ray excess
on the eastern part of the cloud. The area where the brightening occurs is
consistent with that found in the {\em Chandra} data. 
Using regions 1--3 for the X-ray bright area
and 6--11 as off-cloud, low \ion{H}{1} area, we derive average fluxes of
$2.54\pm0.09$ and $1.90\pm0.05$ cts ksec$^{-1}$ arcmin$^{-2}$ or a
difference of $0.64\pm0.10$ cts ksec$^{-1}$ arcmin$^{-2}$. This
corresponds to a 6.4$\sigma$ brightening, which is larger than the 1.6$\sigma$
increase measured in the {\em Chandra} data. The main reason for the difference
between the two data sets is the field of view with the {\em XMM\/}-PN data
(especially regions 6--11) reaching the edge of the cloud, whereas the
{\em Chandra} field of view is situated fully on MS30.7.

The detection limit for point sources in the 0.5--2.0 keV band of our
observation is $2.25\times10^{-15}$ ergs s$^{-1}$ cm$^{-2}$. Using the
distribution of sources by \citet{more03}, we derive a fraction of unresolved
sources for the background of 28\%, corresponding to $0.21$ cts
ksec$^{-1}$ arcmin$^{-2}$ in the 0.4--1.0 keV band. The X-ray brightening of the
Magellanic Stream cloud could therefore be as high as $0.85\pm0.10$
cts ksec$^{-1}$ arcmin$^{-2}$. The 3$\sigma$ upper limit for a shadow based on
regions 6--11 corresponds to 8\% of the off-cloud background, which is
consistent with the range derived from the {\em Chandra} data.

\bigskip

\section{Interpretation of the Results}

\subsection{The Absence of X-Ray Shadows}

Our observations failed to show a shadow cast by the higher column density
regions, although the observation for NGC 5907 produced an order of magnitude 
less data than expected, so this observation does not provide strong
constraints. For our two best cases, MS30.7 and CHVC 125, the pointings toward
the higher \ion{H}{1}
column regions showed a slight brightening rather than a shadow (Table 4).
For CHVC 125, this brightening is only at the 1.6$\sigma$ level, and 
the uncertainty in the ratio of the two pointings (on-cloud and
off-cloud) is 8.5\%, or a 3$\sigma$ upper limit to the shadow of 25\%.  
For MS30.7, the observations with {\it Chandra\/} and {\it XMM-Newton\/} are
consistent in that the flux differences are the values expected from the
difference in instrumental sensitivities.  The difference in the ratios of 
the on-cloud and off-cloud fluxes is about at the 2$\sigma$ level, and if not
merely due to statistical fluctuations, this may be caused by the slightly
different regions used by the two instruments.  The field of view is larger for
the EPIC camera on {\it XMM-Newton\/} than for the ACIS S3 chip on
{\it Chandra\/}, so it was possible to take off-cloud regions further from
on-cloud regions, which may have led to the greater contrast. In these
observations, the flux is brighter for the on-cloud pointings, with a S/N of 1.7
for the {\it Chandra\/} measurements and 7 for the {\it XMM-Newton\/}
data.  When combined by weighting, the brightening is about a 7$\sigma$ feature.
This could occur if there is an interaction between this Magellanic Stream
cloud and a hot diffuse medium.  Ignoring the brightening, the 3$\sigma$
detection limit is 14\% of the diffuse background.

In forming these ratios and upper limits from the on and off-cloud pointings,
we have not removed the contribution from the Local Hot Bubble, which is
certainly not being shadowed by the objects that we pointed at.
We will assume that the contribution from the Local Hot Bubble is the same
in the on and off-cloud locations, in which case, that contribution must be
estimated.

To estimate this contribution from the Local Hot Bubble in the 0.1--0.28 keV
region (ROSAT bands 1+2) we assume it is equal to that of the contribution above
the disk, which is similar to values discussed in the literature \citep{snow00}.
There is significant variation in this ratio around the sky and there are
regions where the halo contribution dominates that from the LHB and regions
where the reverse is true.  The field for the CHVC ({\it l, b\/} = 125\degr ,
41\degr) is in a fairly typical region for the northern Galactic hemisphere,
where the 1/4 keV background is intermediate in intensity between the bright
regions near the poles (the ROSAT measurements of the background at this
location are R12 = 638 and R45 = 137 in units of 10$^{-6}$ cts s$^{-1}$
arcmin$^{-2}$).  Separation between the Local Hot Bubble and emission above the
disk can be made only on statistical grounds
in general and \citet{snow00} show that in the Northern Hemisphere, the two
components are comparable.  For the field of MS30.7, which is very close
to the South Galactic Pole ({\it l, b\/} = 30\degr , -82\degr), the background
is very similar to the CHVC 125 (R12 = 605 and R45 = 128 in units of 10$^{-6}$
cts s$^{-1}$ arcmin$^{-2}$), so we assume the same ratio 
of emission from the LHB to that above the plane.

The 1/4 keV emission from above the disk is absorbed by the disk gas,
but this absorption decreases with energy so that in the 0.4--1.0 keV band,
the emission from above the disk becomes relatively greater.  Assuming
that the temperature and abundances of the two components are the same
(about 0.1 keV), 70--75\% of the diffuse 0.4--1.0 keV background originates
above the disk.  If the component above the disk is hotter as well,
the fraction is greater.
When we assume that 25\% of the diffuse emission in the 0.4--1.0 keV band 
is due to the Local Hot Bubble, there is an increased fractional error (by 33\%) 
in the remaining flux, so the 3$\sigma$ limits for MS30.7 and CHVC 125 
rise to 18\% and 34\% (Table \ref{shadows}).

For CHVC 125, the mean difference between the higher and lower \ion{H}{1}
column density regions is 5$\times$10$^{20}$ cm$^{-2}$ versus
3$\times$10$^{20}$ cm$^{-2}$.  If most of the emission above the disk lay beyond
the CHVC and its gas temperature is
0.1 keV, the shadow would produce a deficit of 17.1\% (for a temperature
of 0.2 keV, that reduces to 14.3\%).  This would have been about a 1.5-2$\sigma$
shadow rather than the observed 1.7$\sigma$ brightening.
The failure to see a shadow for this object does not place strong constraints
on the nature of the Galactic Halo or Local Group medium.  
Since this CHVC probably lies within 100 kpc of the Milky Way \citep{putm03b,malo03,pisa04},
it could interact with a hot Galactic Halo to produce X-ray emission that could
exceed the deficit expected from a shadow.  In addition to the marginal brightening
of X-rays toward the cloud, there is an X-ray enhancement just east of the cloud
edge that may signal an interaction of the neutral cloud with a hot ambient medium.
This is not the first time that a modest X-ray brightening has been seen toward or near
a HVC, as such associations have been reported upon by \citet{herb95}, \citet{kerp96}, and
\citet{kerp99}.

The shadowing constraints are a bit stronger for MS30.7, where the fractional uncertainty
is 6\%.  The difference between the higher and lower \ion{column density H}{1} regions is
4$\times$10$^{20}$ cm$^{-2}$ compared to 2.5$\times$10$^{20}$ cm$^{-2}$, 
and for emission from a 0.1 keV thermal plasma, the shadow depth would 
be 13.1\%, or a predicted 2.2$\sigma$ shadow.
In contrast, we find a brightening above the 7$\sigma$ level, so the emission
from the cloud obscures the predicted shadow, should one exist.  The most likely
explanation for the brightening is that this cloud, at a distance
of 50--70 kpc, is interacting with a hot ambient environment (see below).

The failure to detect a shadow of the diffuse cosmic X-ray
background in the {\it Chandra\/} data does not place a very useful upper limit
on this component.  The limit is that less than 50\% of the diffuse X-ray
background comes from beyond MS30.7 and CHVC 125.  
However, if one could detect excess X-ray emission from several clouds
with known distances around the Milky Way, it might be possible to probe the
density of this hot Galactic halo.

The observation toward NGC 5907 was designed to determine the fraction of the 
X-ray background that could be due to a cosmological diffuse component arising
from the warm-hot intergalactic medium (WHIM).  The surface brightness of the
WHIM is model-dependent, but it could be as large as 20\% of the soft X-ray background in
the 0.4--1.0 keV range \citep{cen99,croft01}.  Unfortunately, the measurement
through the \ion{H}{1} of NGC 5907 only yielded a 1$\sigma$ rms 
of 17\%, so this observation does not provide an important constraint.  If the X-ray background from the WHIM is near 20\%, then an
observation that is an order of magnitude longer (the original
observing request), the rms would drop to 6\% and a shadow would be seen at the
3$\sigma$ level.  However, if the emission from the WHIM is closer to 5\% of the soft X-ray background, the time required to detect the shadow would exceed 1 Msec with current instruments.
\subsection{The X-Ray Enhancement Around MS30.7-81.4-118}

One of the positive outcomes of this work is the detection of an X-ray
enhancement at the location of the Magellanic Stream cloud MS30.7-81.4-118. 
From the measurement with EPIC, the net excess flux toward the cloud is
9${\times}10^{-15}$ erg cm$^{-2}$ s$^{-1}$, for an assumed thermal spectrum with kT = 0.3 keV
and 20\% solar metallicity (the flux is not strongly dependent on the
metallicity).  For a distance to the cloud of 60 kpc, the implied luminosity is
4${\times}10^{33}$ erg s$^{-1}$.  Also, the effective diameter of the cloud is about 70 pc, but
it is not spherically symmetric.

This cloud presumably sits in the ambient medium that surrounds the Milky
Way, but whose parameters are poorly known.  The density of the ambient
medium 60 kpc from the Milky Way is likely to be near $\sim$ 10$^{-4}$ cm$^{-3}$, and it
cannot be larger than a few times this value or it would exceed some of the
dispersion measures toward Large Magellanic Cloud (LMC) pulsars \citep{manch2006}. 
Also, significantly larger
electron densities along 60 kpc path lengths are ruled out by both the OVII
absorption line measurement and the diffuse soft X-ray emission \citep{breg07}.  
A lower limit to the mean ambient density is given from an argument involving the
ram-pressure stripping of material within dwarf spheroidals in the Local Group \citep{blitz00,grce07}.
\citet{blitz00} argue that within 250 kpc, the mean gas density must be at least 
2.5$\times$10$^{-5}$ cm$^{-3}$.
The temperature of this gas is likely to be about 3${\times}10^{6}$ K, if it is near
virial equilibrium with the potential well of the Milky Way and the Local Group.  This would
imply that the ambient pressure is P/k $\sim$ 10$^{2.5} (n/10^{-4} cm^{-3})$ K cm$^{-3}$.

In contrast, if the emission enhancement detected by XMM comes from a
uniform density region with the above properties, the electron density is
about 4$\times$10$^{-3}$ cm$^{-3}$, which would 
imply a pressure of P/k = 2$\times$10$^{4}$ K cm$^{-3}$. 
This pressure is nearly two orders of magnitude larger than that of the ambient medium.

A likely mechanism for the production of this emission is the differential
motion between the HI cloud and the hot ambient medium.  This
differential velocity is about 300 km s$^{-1}$ \citep{vdMar2002}, 
although recent observations show that it may be closer to 370 km s$^{-1}$
\citep{kall06,besla07}.  To be conservative, we will assume that the hot ambient
medium is at rest with respect to the Milky Way and that the Magellanic
Stream is moving through it at 300 km s$^{-1}$.  As a crude measure of
the energy available, we estimate the ram pressure heating rate by the
expression $\frac{\pi}{2}\rho r_{c}^{2}v^{3}$, where $r_{c}$ is the cloud radius,
$\rho$ is the ambient gas density and $v$ is the relative velocity.  
Using this, we find the power available is 1${\times}10^{35}(n/10^{-4} cm^{-3})$ erg s$^{-1}$.  
This is about a factor of twenty larger than the observed power, 
so there is an adequate power supply to account for the observed 
radiative emission, even if the density is 2.5$\times$10$^{-5}$ cm$^{-3}$.  
We now consider how this energy might be transferred from relative 
motion to thermal energy.

This cloud motion might drive a weak shock into the ambient medium,
since the differential velocity is comparable to the sound speed.  Even if it
were a strong shock, it could only raise the density by a factor of four,
rather than the factor of about 40 that is observed.  A more likely explanation is
that the emission occurs from the shock that is driven into the neutral cloud,
heating the gas and causing this overpressure medium to flow away from
the HI cloud.  Additional processes are likely to be important, such as
mixing between the HI cloud and the hot medium \citep{bland07,parr08}.

These energy budget and plausibility arguments are helpful in understanding
the processes that lead to X-ray emission, but they are no substitute for a
serious simulation.  Such a numerical hydrodynamic simulation was carried 
out by \citet{bland07} for the purpose of modeling the optical H$\alpha$ from
the Magellanic Stream clouds.  In the future, it may be possible to extend
these calculations to understand whether the interaction between a Magellanic 
Stream cloud and a hot ambient medium can cause the observed X-ray brightening.

\acknowledgements
We are especially grateful to Robert Braun and Butler Burton for making the
\ion{H}{1} total intensity map of CHVC 125+21-207 available to us.  Also, we
would like to thank Renato Dupke, Chris Mullis, and Steve Snowden for their
suggestions and assistance and to the {\it Chandra\/} and {\it XMM-Newton\/}
teams for answering many of our questions.  We gratefully acknowledge financial
support for this research, which was provided by NASA.

\clearpage
\begin{deluxetable}{ccrrrrr}
\tabletypesize{\scriptsize}
\tablecaption{X-Ray Observations}
\tablewidth{0pt}
\tablehead{
\colhead{Target} & \colhead{Instrument} & \colhead{ObsID} & \colhead{RA (J2000)}
& \colhead{DEC (J2000)} & \colhead{t$_{exp}$ (ksec)} &
\colhead{t$_{clean}$ (ksec)} 
 }
\startdata
NGC 891 & Chandra & 794 & 02 22 33.4 & 42 20 57.0 & 51.56 & 36.15 \\ 
NGC 891 & Chandra & 4613 & 02 22 31.3 & 42 19 57.3 & 120.4 & 108.5 \\ 
CHVC125 & Chandra & 2253 & 12 29 10.0 & 75 23 45.0 & 48.01 & 45.27 \\ 
CHVC125 & Chandra & 2484 & 12 28 20.0 & 75 23 45.0 & 24.34 & 10.09 \\ 
CHVC125 & Chandra & 2486 & 12 27 30.0 & 75 23 45.0 & 22.96 & 21.2 \\ 
MS30.7 & Chandra & 5038 & 00 13 18.9 & $-$27 13 24.0 & 50.14 & 49.51 \\ 
MS30.7\_off & Chandra & 5039 & 00 15 05.5 & $-$27 27 34.7 & 49.88 & 33.49 \\ 
MS30.7 & XMM & 204670101 & 00 12 56.3 & $-$27 12 06.7 & 50.03 & 37.0 \\ 
NGC 5907 & XMM & 145190101 & 15 15 59.0 & 56 20 34.4 & 52.19 & 5.9 \\ 
NGC 5907 & XMM & 145190201 & 15 15 58.9 & 56 20 33.7 & 52.74 & 6.6 \\ 
\enddata
\end{deluxetable}

\clearpage

\begin{deluxetable}{lcccc}
\tablewidth{0pt}
\tablecaption{\label{n59meas} NGC\,5907 Measurements}
\tablehead{\colhead{Method} & \colhead{Region} & \colhead{On-galaxy} &
\colhead{Off-galaxy} & \colhead{On--Off} \\
\colhead{} & \colhead{} & \colhead{(cts ksec$^{-1}$ arcmin$^{-2}$)} &
\colhead{(cts ksec$^{-1}$ arcmin$^{-2}$)} & \colhead{(cts ksec$^{-1}$
arcmin$^{-2}$)}}
\startdata
$\chi^2$ & 1 & $1.12\pm0.13$ & $1.57\pm0.16$ & $-0.45\pm0.21$ \\
$\chi^2$ & 2 & $1.78\pm0.17$ & $1.49\pm0.16$ & $+0.29\pm0.23$ \\
CFW      & 1 & $1.56\pm0.23$ & $2.15\pm0.26$ & $-0.60\pm0.34$ \\
CFW      & 2 & $2.36\pm0.28$ & $1.99\pm0.26$ & $+0.34\pm0.39$ \\
\enddata
\end{deluxetable}

\clearpage

\begin{deluxetable}{ccc}
\tablewidth{0pt}
\tablecaption{\label{mscmeas} MS30.7 Measurements}
\tablehead{\colhead{Region} & \colhead{Flux} & On--Off Flux \\
\colhead{} & \colhead{(cts ksec$^{-1}$ arcmin$^{-2}$)} & \colhead{(cts 
ksec$^{-1}$ arcmin$^{-2}$)}}
\startdata
 1 & $  2.39\pm  0.15$ & $  0.69\pm  0.15$ \\
 2 & $  2.66\pm  0.15$ & $  0.96\pm  0.15$ \\
 3 & $  2.56\pm  0.15$ & $  0.85\pm  0.15$ \\
 4 & $  2.15\pm  0.14$ & $  0.45\pm  0.15$ \\
 5 & $  2.27\pm  0.14$ & $  0.56\pm  0.15$ \\
 6 & $  1.71\pm  0.12$ & $  0.00\pm  0.13$ \\
 7 & $  1.95\pm  0.13$ & $  0.25\pm  0.14$ \\
 8 & $  2.32\pm  0.14$ & $  0.61\pm  0.15$ \\
 9 & $  1.79\pm  0.13$ & $  0.09\pm  0.14$ \\
10 & $  1.76\pm  0.12$ & $  0.05\pm  0.14$ \\
11 & $  1.84\pm  0.13$ & $  0.13\pm  0.15$ \\
12 & $  2.28\pm  0.15$ & $  0.57\pm  0.16$ \\
13 & $  2.10\pm  0.13$ & $  0.39\pm  0.14$ \\
14 & $  2.09\pm  0.13$ & $  0.38\pm  0.14$ \\
15 & $  1.90\pm  0.12$ & $  0.19\pm  0.14$ \\
16 & $  1.85\pm  0.12$ & $  0.14\pm  0.14$ \\
17 & $  1.92\pm  0.13$ & $  0.21\pm  0.14$ \\
18 & $  1.92\pm  0.13$ & $  0.21\pm  0.15$ \\
19 & $  1.89\pm  0.13$ & $  0.18\pm  0.14$ \\
20 & $  1.96\pm  0.14$ & $  0.25\pm  0.15$ \\
\enddata
\end{deluxetable}

\clearpage

\begin{deluxetable}{lcllccccccc}
\tabletypesize{\scriptsize}
\tablewidth{0pt}
\rotate
\tablecaption{\label{shadows} Surface Brightness Contrasts}
\tablehead{
\colhead{Object} & \colhead{Observatory} & \colhead{On Cloud} &
 \colhead{Off Cloud} & \colhead{Instrumental} & \colhead{Unresolved} &
 \colhead{On} & \colhead{Off} & \colhead{Ratio1} & \colhead{Local} &
 \colhead{Ratio2} \\
\colhead{} & \colhead{} & \colhead{(raw)} & \colhead{(raw)} &
 \colhead{Background} & \colhead{Pt. Src.} & \colhead{(net)} & \colhead{(net)} &
 \colhead{} & \colhead{Bubble}
}
\startdata
MS30.7 & Chandra & $ 1.29\pm 0.042$ & $ 1.20\pm 0.04$ & 0.61 & 0.04 & 0.64
 & 0.55 & $ 1.16\pm 0.097$ & 0.15 & $ 1.22\pm  0.130$ \\
CHVC 125 & Chandra & $ 1.20\pm  0.033$ & $ 1.14\pm  0.031$ & 0.61 & 0.03 & 0.56
 & 0.50 & $ 1.12\pm  0.085$ & 0.13 & $ 1.16\pm  0.114$ \\
MS30.7 & XMM & $ 3.24\pm  0.09$ & $ 2.60\pm  0.05$ & 0.70 & 0.21 & 2.33
 & 1.69 & $1.38\pm  0.051$ & 0.50 & $1.54\pm  0.068$ \\
NGC 5907 & XMM & $ 1.90\pm  0.11$ & $ 1.98\pm  0.11$ & 0.45 & 0.25 & 1.20
 & 1.28 & $ 0.94\pm  0.125$ & 0.31 & $0.92\pm  0.167$ \\
MS30.7 & Sum & & & & & & & $ 1.33\pm  0.045$ & & $ 1.47\pm  0.060$ \\
\enddata
\tablecomments{
The units of surface brightness are cts ksec$^{-1}$ arcmin$^{-2}$, which apply
to columns 3--8, and 10.  The surface brightness due to the instrumental
background (column 5) and unresolved point sources (column 6) are removed from
the raw surface brightness to produce the net counts (columns 7, 8).  
The quantity Ratio1 is the ratio of the net counts on divided by the net counts
off the cloud (column 7 divided by column 8).  The quantity Ratio2 (column 11)
is the same ratio, after the removal of the surface brightness contribution from
the Local Hot Bubble (column 10).  The sum of the MS30.7 results is a weighted average.
}
\end{deluxetable}

\clearpage

\end{document}